\documentclass[runningheads]{llncs}
\usepackage[T1]{fontenc} 
\usepackage{multirow}
\usepackage{graphicx}
\usepackage{xspace}
\usepackage{algorithm}
\usepackage{url}
\usepackage{subfig}
\usepackage[normalem]{ulem}
\usepackage{algpseudocode}
\newcommand{\modelname}{PDC-FRS\xspace} 

\begin{document}
\title{\modelname: Privacy-preserving Data Contribution for Federated Recommender System}
\titlerunning{\modelname}
\author{Chaoqun Yang\inst{1} \and
Wei Yuan\inst{2} \and
Liang Qu\inst{2} \and Thanh Tam Nguyen\inst{1}}
\authorrunning{Yang et al.}
% \authorrunning{Anonymity et al.}
% First names are abbreviated in the running head.
% If there are more than two authors, 'et al.' is used.
%
\institute{Griffith University, Gold Coast, Australia 
\\\email{t.nguyen19@griffith.edu.au}\\\and
The University of Queensland, Brisbane, Australia}

% If the paper title is too long for the running head, you can set
% an abbreviated paper title here
%
% \author{Chaoqun Yang\inst{1} \and
% Wei Yuan\inst{2} \and
% Liang Qu\inst{2} \and
% Thanh Tam Nguyen\inst{1}
%
% \authorrunning{Yang et al.}
% \authorrunning{Anonymity et al.}
% First names are abbreviated in the running head.
% If there are more than two authors, 'et al.' is used.
%
% \institute{Princeton University, Princeton NJ 08544, USA \and
% Springer Heidelberg, Tiergartenstr. 17, 69121 Heidelberg, Germany
% \email{lncs@springer.com}\\
% \url{http://www.springer.com/gp/computer-science/lncs} \and
% ABC Institute, Rupert-Karls-University Heidelberg, Heidelberg, Germany\\
% \email{\{abc,lncs\}@uni-heidelberg.de}}
%
\maketitle              % typeset the header of the contribution
\begin{abstract}
Federated recommender systems (FedRecs) have emerged as a popular research direction for protecting users' privacy in on-device recommendations. 
In FedRecs, users keep their data locally and only contribute their local collaborative information by uploading model parameters to a central server. 
While this rigid framework protects users' raw data during training, it severely compromises the recommendation model's performance due to the following reasons: 
(1) Due to the power law distribution nature of user behavior data, individual users have few data points to train a recommendation model, resulting in uploaded model updates that may be far from optimal; 
(2) As each user's uploaded parameters are learned from local data, which lacks global collaborative information, relying solely on parameter aggregation methods such as FedAvg to fuse global collaborative information may be suboptimal.
To bridge this performance gap, we propose a novel federated recommendation framework, \modelname. 
Specifically, we design a privacy-preserving data contribution mechanism that allows users to share their data with a differential privacy guarantee. 
Based on the shared but perturbed data, an auxiliary model is trained in parallel with the original federated recommendation process. 
This auxiliary model enhances FedRec by augmenting each user's local dataset and integrating global collaborative information. 
To demonstrate the effectiveness of \modelname, we conduct extensive experiments on two widely used recommendation datasets. 
The empirical results showcase the superiority of \modelname compared to baseline methods.

\end{abstract}
\section{Introduction}\label{sec_introduction}
Nowadays, the demand for recommender systems has dramatically increased in most online services (e.g., e-commerce~\cite{chen2024adversarial} and social media~\cite{yin2016spatio}) due to their success in alleviating information overload. The remarkable personal recommendation ability of these systems mainly originates from discovering and mining latent user-item relationships from massive user data, such as user profiles and historical behaviors~\cite{lu2015recommender}. Conventionally, this personal user data is collected on a central server, and a recommendation model is trained using it~\cite{he2020lightgcn}, which carries high risks of privacy leakage. With growing awareness of privacy and the recent release of privacy protection regulations, these traditional centralized recommender model training paradigms have become harder to implement~\cite{yang2020federated}.

Federated learning~\cite{zhang2021survey} is a privacy-preserving decentralized training scheme that collaboratively trains a model without sharing clients' sensitive raw data. Consequently, many works investigate the combination of federated learning and recommender systems, known as federated recommender systems (FedRecs)~\cite{ammad2019federated,yin2024device}. In FedRecs, users/clients\footnote{In this paper, we focus on cross-user federated recommendation, where each client represents one user. Therefore, the concepts of user and client are interchangeable.} manage their own data locally while a central server coordinates the training process. Specifically, the central server distributes recommendation model parameters to clients. Upon receiving the recommendation model, clients train it on their local data and send the updated model parameters back to the central server. The central server then aggregates these parameters to integrate each client's collaborative information.

Although the generic federated recommendation framework conceals users' raw data and thus protects their privacy, it may only achieve suboptimal performance for the following reasons. Firstly, since user-item interaction data follows a power law distribution~\cite{zaier2008evaluating}, most users have few data points to train a recommendation model effectively. Consequently, the locally trained parameters will be far from optimal when sent to the central server. Furthermore, the only way to integrate collaborative information from different clients is by aggregating the uploaded parameters, which may be less effective.

To address the above problems, recent works have broken the rigid FedRec learning protocol by allowing users to control the portion of data they want to share~\cite{anelli2021federank,qu2024towards}. 
In this approach, users can contribute their collaborative information not only by uploading the locally trained parameters but also by sharing their raw data. Unfortunately, we argue that these methods pose high privacy risks since users directly disclose their real and clean data to the central server. For example, when the exposed data is sufficient, a curious central server can easily infer users' remaining data using graph completion techniques~\cite{chen2022review}.

Local differential privacy (LDP) has become the gold standard for providing protection guarantees in federated recommender systems~\cite{li2020federated,wang2022fast,zhang2023comprehensive}. These works apply LDP by transforming users' local model parameters into a noisy version before uploading them to the central server. Inspired by this idea, we attempt to leverage LDP to provide a privacy guarantee for user data contributions. Nevertheless, there are significant challenges in achieving differentially private user data contributions. Firstly, current LDP methods in FedRecs are not directly applicable to user data since they are not typical numerical values like model parameters. Additionally, how to effectively leverage perturbed user data in the FedRec learning process remains underexplored.

In this paper, we propose \modelname, the first federated recommendation framework that enables user data contribution with privacy guarantees. Specifically, \modelname utilizes the exponential mechanism to achieve LDP on user-shared data. Then, \modelname simultaneously trains an auxiliary model on the users' uploaded data. This auxiliary model enhances FedRecs in two ways. First, it acts as a data augmentor to enrich each client's local dataset. Additionally, the auxiliary model parameters are fused into FedRecs via contrastive learning to provide global collaborative information. To demonstrate the effectiveness of \modelname, we conduct extensive experiments on two datasets. The experimental results indicate the superiority of our method compared to several federated recommendation baselines.

To sum up, the main contributions of this paper are as follows:
\begin{itemize}
    \item[$\bullet$] To the best of our knowledge, we are the first to investigate user data contribution with a privacy-preserving mechanism in a federated recommender system (\modelname), allowing users to share their data with differential privacy guarantees.
    \item[$\bullet$] In \modelname, we design an exponential mechanism-based privacy protection for user data and provide two methods for using the perturbed data to enhance federated recommendation performance.
    \item[$\bullet$] We conduct extensive experiments on public real-world recommendation datasets to validate the effectiveness of \modelname. The experimental results demonstrate the promising performance of our proposed methods.
\end{itemize}

\section{Related Work}
\subsection{Centralized Recommender System}
Recommender systems aim to provide personalized recommendations to users based on their historical data (e.g., purchases and browses)~\cite{lu2015recommender,zheng2023automl}.
Generally, these systems can be classified into matrix factorization (MF)-based methods~\cite{koren2009matrix,chen2022review}, deep learning-based methods~\cite{covington2016deep,he2017neural}, and graph neural network (GNN)-based methods~\cite{wang2019neural,he2020lightgcn}.
MF methods decompose the user-item rating matrix into two lower-dimensional vectors representing user and item preferences. Deep learning-based systems utilize deep learning models to discover nonlinear relationships between users and items. Recently, GNN-based recommendation has gained popularity as GNN models can effectively capture high-order user-item interactions, treating user-item interactions as a graph structure. 
However, all these methods are traditionally trained in a centralized manner, which raises privacy concerns.

\subsection{Federated Recommender System}
Federated recommender systems (FedRecs) have garnered significant attention recently due to their privacy protection capabilities~\cite{yin2024device,Wang2024PoisoningAA}.
Ammad et al.~\cite{ammad2019federated} were pioneers in designing the first federated recommendation framework, spawning subsequent research in this area~\cite{sun2022survey}.
These studies can be categorized into four perspectives based on their objectives: effectiveness~\cite{wang2022fast,yuan2023hetefedrec,luo2022personalized,yuan2024robust}, efficiency~\cite{liang2021fedrec++,muhammad2020fedfast}, privacy~\cite{yuan2023federated,yuan2023interaction,zhang2023comprehensive}, and security~\cite{zhang2022pipattack,yuan2023manipulating,yuan2023manipulating1}.
For instance, Wu et al.~\cite{wu2021fedgnn} and Yuan et al.~\cite{yuan2024robust} applied advanced neural networks and contrastive learning in FedRecs to enhance recommendation performance. 
Muhammad et al.~\cite{muhammad2020fedfast} investigated efficient training methods for FedRecs with rapid convergence, while Zhang et al.~\cite{zhang2023comprehensive} explored sensitive attribute privacy in FedRecs. 
However, these works operate within the original federated recommendation framework where users can only contribute collaborative information by uploading model parameters, which may limit effectiveness.
Anelli et al.~\cite{anelli2021federank} and Qu et al.~\cite{qu2024towards} highlighted this limitation and proposed FedRecs frameworks based on user-governed data contribution. 
Nonetheless, in their approaches, parts of users' clean data are directly transmitted to the central server, lacking privacy guarantees and posing significant privacy risks. 
In response, our paper introduces a privacy-preserving user data contribution framework for FedRecs.

\begin{figure}[!t]
\centering
    \includegraphics[width=0.8\textwidth]{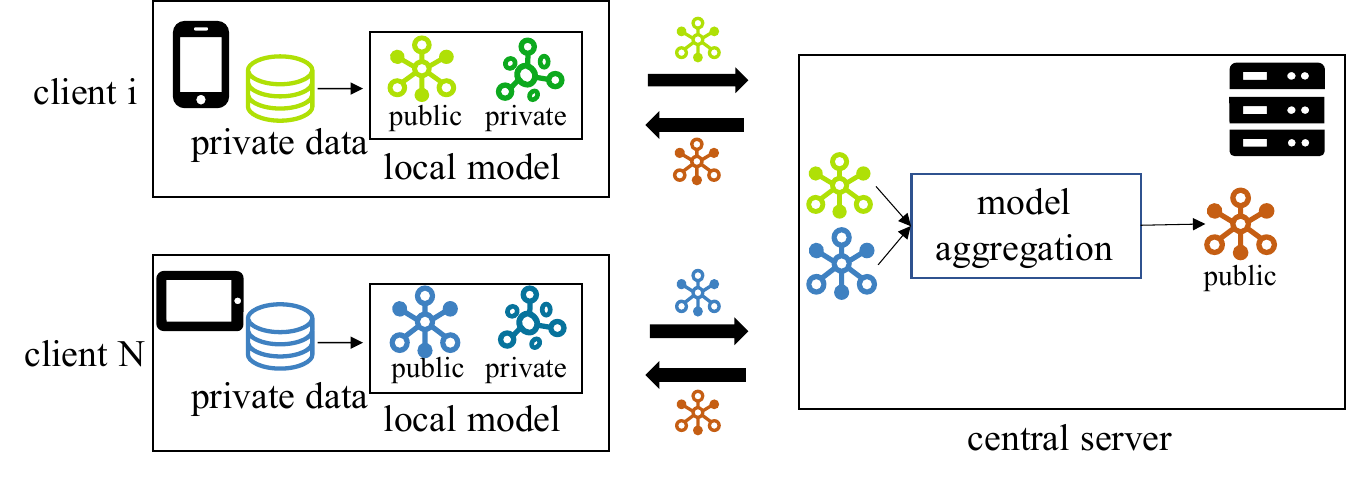}
    \caption{Overview of a typical federated recommender system} \label{fig_fedrec}
\end{figure}

\section{Preliminaries}\label{sec_preliminaries}
\subsection{Federated Recommendation Framework}
Let $\mathcal{U} = \{u_{i} \}_{i=1}^{\left|\mathcal{U}\right|}$ and $\mathcal{V} = \{v_{j}\}_{j=1}^{\left|\mathcal{V}\right|}$ represent the set of users and items in FedRec, where $\left|\mathcal{U}\right|$ and $\left|\mathcal{V}\right|$ are the number of users and items, respectively.
$\mathcal{D}_{u_{i}}$ is user $u_{i}$'s raw private dataset, which contains a set of triples $(u_{i}, v_{j}, r_{ij})$ representing its interaction status.
$r_{ij}=1$ indicates $u_{i}$ has interacted with item $v_{j}$ while $r_{ij}=0$ means that $v_{j}$ is still in non-interacted item pool\footnote{In this paper, we mainly focus on recommendation with implicit feedback.}.
In traditional FedRec, $\mathcal{D}_{u_{i}}$ is always stored in the user's local device and cannot be used by any other participants.
The goal of FedRec is to train a recommender model that can predict user preference tendency for non-interacted items using these distributed private datasets.

To achieve this goal, most FedRec systems follow the learning protocol shown in Figure~\ref{fig_fedrec}.
A central server acts as a coordinator to manage the entire training process, with the recommendation model divided into public and private parameters.
Private parameters typically refer to user embeddings $\mathbf{U}$, while public parameters primarily consist of item embeddings $\mathbf{V}$.
Clients engage in collaborative learning by transmitting public parameters.
Specifically, there are four steps iteratively executed during the FedRec learning process. 
At first, the central server selects a group of clients $\mathcal{U}_{t-1}$ to participate and disperses public parameters $\mathbf{V}^{t-1}$ to them.
Then, the client $u_{i}\in \mathcal{U}_{t-1}$ utilizes the received public parameters $\mathbf{V}^{t-1}$ and its corresponding local parameters $\mathbf{u}_{i}$ to train the recommendation model on their local datasets $\mathcal{D}_{u_{i}}$ using certain objective function, such as:
\begin{equation}\label{eq_naive_recloss}
      \mathcal{L}^{original} = -\sum\nolimits_{(u_{i}, v_{j}, r_{ij})\in \mathcal{D}_{u_i}} r_{ij}\log \hat{r}_{ij} + (1-r_{ij})\log (1-\hat{r}_{ij})
  \end{equation}
After training, the clients send the updated public parameters $\mathbf{V}^{t-1}_{u_{i}}$ back to the central server.
Finally, the central server aggregates these uploaded parameters $\{\mathbf{V}^{t-1}_{u_{i}}\}_{u_{i}\in \mathcal{U}_{t-1}}$ to get a new version of public parameters $\mathbf{V}^{t}$.
It is evident that clients' collaborative information can only be conveyed through the public parameters $\mathbf{V}^{t-1}_{u_{i}}$ via aggregation, which may be less effective.

\subsection{Base Recommendation Model}
Federated recommendation framework is compatible with most deep learning based recommender systems.
In this paper, without loss of generality, we adopt neural collaborative filtering (NCF)~\cite{he2017neural} as the experimental base models, which are widely used in this research area~\cite{ammad2019federated,yuan2023hetefedrec,zhang2022pipattack}.
Based on matrix factorization, NCF leverages a multi-layer perceptron (MLP) and the concatenation of user and item embeddings to predict a user's preference score for an item:
\begin{equation}
    \hat{r}_{ij} = \sigma (\mathbf{h}^{\top}MLP([\mathbf{u}_{i}, \mathbf{v}_{j}]))
\end{equation}

\begin{figure}[!t]
\centering
    \includegraphics[width=0.8\textwidth]{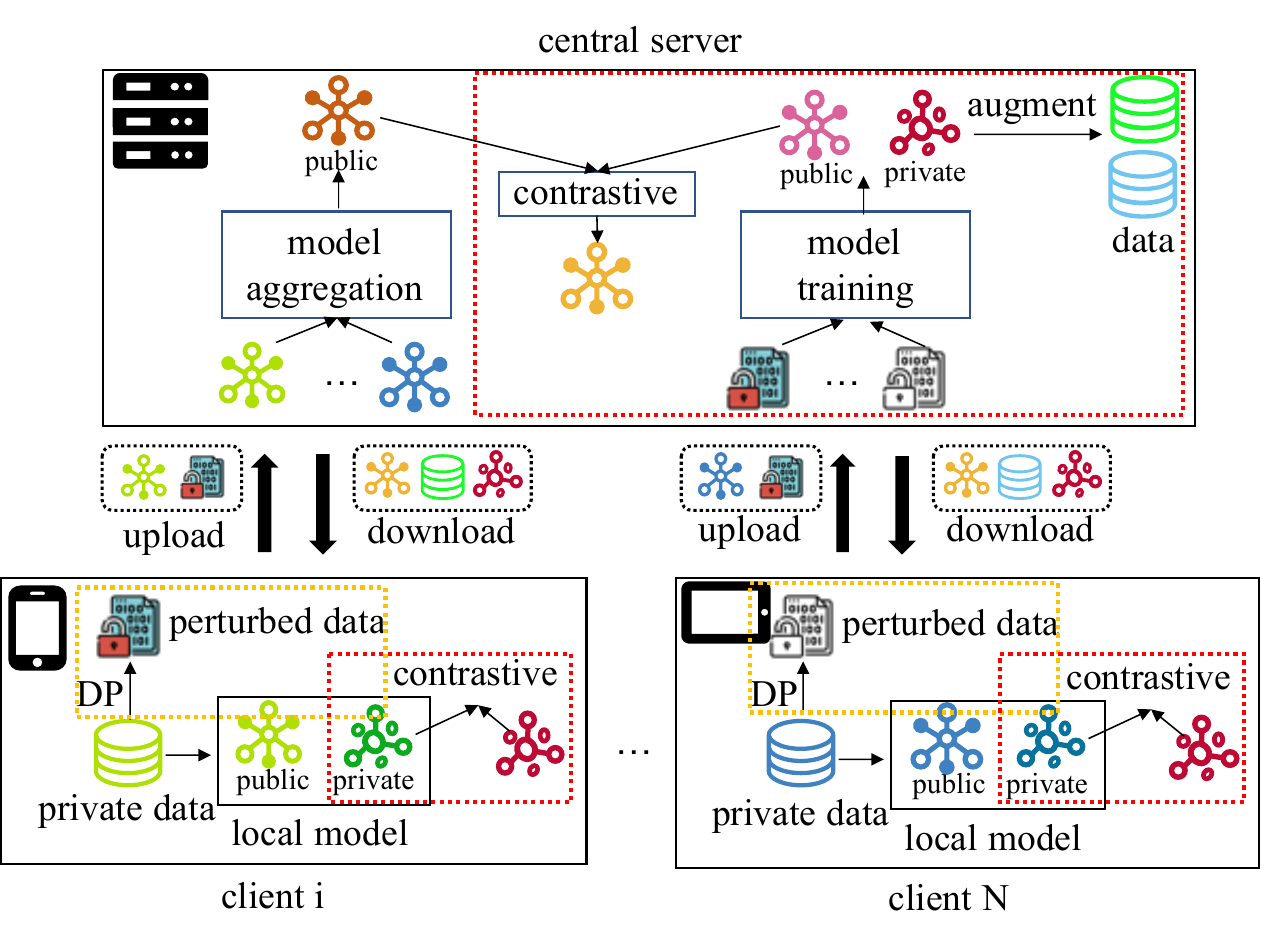}
    \caption{Overview of \modelname. The yellow dotted box highlights the privacy-preserving data contribution while the red dotted box depicts the methods of using data contribution to enhance federated recommendation, which are the two most important parts of \modelname.} \label{fig_overview}
\end{figure}

\section{Methodology}
In this section, we present the technical details of our \modelname. We first introduce the general framework of \modelname in Section~\ref{sec_pdc_overview}, and then discuss the specific component designs. Specifically, we describe how to achieve privacy-preserving data contribution in Section~\ref{sec_privacy_data_contribution}. In Section~\ref{sec_enhance_with_data_contribution}, we introduce the details of utilizing an auxiliary model to enhance FedRec performance.

\subsection{General Framework of \modelname}\label{sec_pdc_overview}
The traditional federated recommender system depicted in Figure~\ref{fig_fedrec} relies solely on public parameters to convey clients' collaborative information. This approach has limitations because (1) each client's uploaded parameters vary in quality and (2) the uploaded parameters are trained only on local data, lacking global information.

In this work, we propose \modelname to enhance FedRec's performance by allowing users to contribute to the system using data in a privacy-preserving manner. Revealing users' clean data to the central server, as done in~\cite{anelli2021federank,qu2024towards}, would significantly compromise FedRec's privacy-preserving capability. Therefore, the first challenge for \modelname is ``how to share user data in a privacy-preserving manner?"
As shown in Figure~\ref{fig_overview}, we employ a differential privacy method to transform users' private data into a noisy version before uploading it. However, this perturbation reduces the utility of the uploaded data, presenting another challenge for \modelname: ``how to leverage these noisy data to enhance federated recommendation performance?"
In this paper, we first train an auxiliary model on the user-uploaded data and then explore two ways to enhance FedRec: data augmentation and contrastive learning. In the following two subsections, we detail how we address these two challenges.

\subsection{Exponential Mechanism-based Data Contribution}\label{sec_privacy_data_contribution}
Local Differential Privacy (LDP)~\cite{arachchige2019local,zhang2023trajectory}, a variant of Differential Privacy (DP)~\cite{mcsherry2007mechanism}, provides a rigorous mathematical privacy guarantee and has thus become the \textit{de facto} standard for privacy protection in decentralized settings. 
Consequently, in \modelname, we design a data contribution mechanism that satisfies LDP.

\textit{Definition 1: Local Differential Privacy (LDP).}
Let $\mathcal{A}(\cdot)$ be a randomized function, for example, a perturbation algorithm or machine learning model, and $\epsilon>0$ is the privacy budget.
The function $\mathcal{A}(\cdot)$ is said to provide $\epsilon$-Local different privacy if for any two input data $x_{1}, x_{2} \in \mathcal{X}$ and for any possible output $y\in\mathcal{Y}$ satisified the following inequality:
\begin{equation}\label{eq_dp}
    Pr[\mathcal{A}(x_{1}) = y] \leq exp(\epsilon) Pr[\mathcal{A}(x_{2})=y]
\end{equation}
where $Pr[\cdot]$ is the probability and $\epsilon$ controls the trade-off between data utility and privacy.
Intuitively, E.q.~\ref{eq_dp} implies that based on the output of algorithm $\mathcal{A}(\cdot)$, the adversarial attacker cannot accurately infer the original data is $x_{1}$ or $x_{2}$, since their output probability difference of value $y$ is bounded by $exp(\epsilon)$.

To achieve LDP, each user needs to perturb its protected objectives before sending it to the collector. 
For example, in traditional FedRec, the protected objective is model public parameters $\mathbf{V}$.
A user $u_{i}$ will locally transform $\mathbf{V}$ into a noisy version $\mathbf{V}'$ using a perturbation algorithm such as Laplace mechanism before sending it to the central server~\cite{zhang2023comprehensive}.

In \modelname, the user will upload their private dataset $\mathcal{D}_{u_{i}}=\{(u_{i}, v_{j}, r_{ij})\}_{v_{j}\in\mathcal{V}}$ to the central server.
As we mainly focus on the recommendation with implicit feedback, the actual protection objective is user $u_{i}$'s interacted item set $\mathcal{V}_{u_{i}}^{+}$.
Therefore, before sending data contribution, the user needs to perturb $\mathcal{V}_{u_{i}}^{+}$ to a noisy version $\mathcal{V}_{u_{i}}^{+'}$
As $\mathcal{V}_{u_{i}}^{+}$ is a set, in this paper, we adopt the exponential mechanism as the perturbation algorithm to protect user data.

\textit{Definition 2: Exponential Mechanism.} The exponential mechanism $\mathcal{A}$ is said to preserve $\epsilon$-LDP if for any input $x\in\mathcal{X}$, the probability of any output $y\in \mathcal{Y}$ is:
\begin{equation}
    Pr[\mathcal{A}(x)=y] = \frac{exp(\frac{\epsilon \rho(x,y)}{2\Delta})}{\sum\limits_{y'\in\mathcal{Y}} exp(\frac{\epsilon \rho(x,y')}{2\Delta})}
\end{equation}
where $\epsilon$ is the privacy budget, $\rho(\cdot)$ is the rating function, and $\Delta$ is the sensitivity of $\rho(\cdot)$.

In \modelname, we utilize the similarity between two items $v_{i}$ and $v_{j}$ as the rating function to sample a new item set $\mathcal{V}_{u_{i}}^{+'}$ based on $\mathcal{V}_{u_{i}}^{+}$, which can protect privacy while keep data utility, since replacing $v_{i}$ with similar item $v_{j}$ can still keep user's expected collaborative information to a large extent:
\begin{equation}\label{eq_data_protect_final}
    Pr[\mathcal{A}(v_{i})=v_{j}] = \frac{exp(\frac{\epsilon sim(v_{i},v_{j})}{2\Delta})}{\sum\limits_{v_{k}\in\mathcal{\mathcal{V}}} exp(\frac{\epsilon sim(v_{i},v_{k})}{2\Delta})}
\end{equation}
However, how to calculate the similarity score is not trivial. 
A simple method is to directly compute based on the item embedding table $\mathbf{V}$.
Unfortunately, in the first few rounds, the item embedding table is from scratch and therefore contains less useful information, making the similarity scores low-quality.
Considering that in most cases, items contain content information (e.g., title), and this information can represent items to some extent.
Without loss of generality, in this paper, we calculate the similarity based on item titles using pre-trained word embeddings.

\subsection{Enhance Federated Recommendation with Data Contribution}\label{sec_enhance_with_data_contribution}

After the central server received all users privacy-preserving data contributions $\{\mathcal{D}_{u_{i}}'\}_{u_{i}\in\mathcal{U}}$, the central server constructs an auxiliary model $\mathbf{M}_{aux}=\{\mathbf{U}_{aux}, \mathbf{V}_{aux}\}$ and will train the model on the perturbed dataset.
Specifically, before a global training round $t$ starts, the central server updates  $\mathbf{M}_{aux}^{t-1}$ to $\mathbf{M}_{aux}^{t}$ on dataset $\{\mathcal{D}_{u_{i}}'\}_{u_{i}\in\mathcal{U}}$ using the same learning hyper-parameters as the clients' local training.
Then, the auxiliary model $\mathbf{M}_{aux}^{t}$ is used to improve federated recommendation from two perspectives: data augmentation and contrastive learning.

\subsubsection{Auxiliary Model as Data Augmentor.}
As we mentioned in Section~\ref{sec_introduction}, most clients lack training data to support the effective updating of the recommendation model.
As $\mathbf{M}_{aux}^{t}$ is trained on a massive dataset that contains multiple users' data, $\mathbf{M}_{aux}^{t}$ can learn some broader collaborative information.
Therefore, we utilize $\mathbf{M}_{aux}^{t}$ as the data augmentor to generate an augmented dataset $\widetilde{\mathcal{D}}_{u_{i}}^{t} = \{(u_{i}, v_{j}, \widetilde{r}_{ij}^{t})\}_{v_{j}\in \mathcal{V}}$ for each user:
\begin{equation} \label{eq_naive_aug}
    \widetilde{r}_{ij}^{t} = f_{rec}(\mathbf{M}_{aux}^{t}|(u_{i}, v_{j})) 
\end{equation}
where $f_{rec}$ is the recommendation algorithm.
However, directly transfer $\widetilde{\mathcal{D}}_{u_{i}}^{t}$  is ineffective since (1) the dataset is very large and (2) $\mathbf{M}_{aux}^{t}$ is trained on perturbed data thus not all predictions are helpful.
To solve this problem, we select a subset $\widetilde{\mathcal{V}}_{u_{i}}$ based on the ranking of $\widetilde{r}_{ij}^{t}$.
Intuitively, the higher $\widetilde{r}_{ij}^{t}$ score indicates that the model has higher confidence in predicting the user will prefer the item $v_{j}$.
Finally, each client $u_{i}$'s local training objective is transformed from E.q.~\ref{eq_naive_recloss} to:
\begin{equation}
    \label{eq_aug_recloss}
      \mathcal{L}^{rec} = -\sum\nolimits_{(u_{i}, v_{j}, r_{ij})\in \mathcal{D}_{u_i}\cup\widetilde{\mathcal{D}}^{t}_{u_{i}}} r_{ij}\log \hat{r}_{ij} + (1-r_{ij})\log (1-\hat{r}_{ij})
  \end{equation}

\subsubsection{Auxiliary Model as Contrastive View.}
As discussed in Section~\ref{sec_introduction}, the other problem of FedRec is that users' collaborative information is only fused via parameter aggregation, such as FedAvg.
In \modelname, we enhance the knowledge fusion across users using contrastive learning.
Specifically, since the auxiliary model $\mathbf{M}_{aux}^{t}$ is trained on the broader dataset that is perturbed by replacing items with their similar items in possibility, we treat the $\mathbf{M}_{aux}^{t}$ as contrastive views.
As shown in Figure~\ref{fig_overview}, after aggregation, the central server obtains the public parameter trained by clients $\mathbf{V}^{t}$.
Then, the central server conducts contrastive learning by treating the counterpart item embeddings in $\mathbf{V}^{t}$ and $\mathbf{V}_{aux}^{t}$ as positive views while others as negative views:
\begin{equation}\label{eq_item_con_loss}
    \mathcal{L}^{itemCL} = \beta \sum_{v_{i}\in\mathcal{V}} -log \frac{exp(sim(\mathbf{v}_{i},\mathbf{v}_{i}^{aux})/\tau)}{\sum_{v_{j}\in \mathcal{V}} exp(sim(\mathbf{v}_{i},\mathbf{v}_{j}^{aux})/\tau)}
\end{equation}
where $\beta$ controls the strengths of $\mathcal{L}^{itemCL}$ and $\tau$ is the temperature.

Besides, \modelname sends the auxiliary model's user embeddings $\mathbf{U}_{aux}^{t}$ to clients for user embedding collaborative learning:
\begin{equation}\label{eq_user_con_loss}
    \mathcal{L}^{userCL} = -log \frac{exp(sim(\mathbf{u}_{i},\mathbf{u}_{i}^{aux})/\tau)}{\sum_{u_{j}\in \mathcal{U}} exp(sim(\mathbf{u}_{i},\mathbf{u}_{j}^{aux})/\tau)}
\end{equation}
Therefore, the client $u_{i}$'s local training objective is transformed from E.q.~\ref{eq_aug_recloss} to:
\begin{equation}\label{eq_user_final_loss}
    \mathcal{L} = \mathcal{L}^{rec} + \lambda\mathcal{L}^{userCL}
\end{equation}
where $\lambda$ controls the strengths of $\mathcal{L}^{userCL}$.
Algorithm~\ref{alg_pdc} summarizes \modelname. 

\begin{algorithm}[!ht]
    \renewcommand{\algorithmicrequire}{\textbf{Input:}}
    \renewcommand{\algorithmicensure}{\textbf{Output:}}
    \caption{The pseudo-code for \modelname.} \label{alg_pdc}
    \begin{algorithmic}[1]
      \Require global round $T$; learning rate $lr$, \dots
      \Ensure  well-trained model $\mathbf{M}^{T} = \{\mathbf{U}^{T}, \mathbf{V}^{T}\}$
      \State each client initializes private parameter $\mathbf{u}_{i}^{0}$
      \State client generates perturbed data $\mathcal{D}_{u_{i}}'$ using E.q.~\ref{eq_data_protect_final} and uploads to the server
      \State server initializes model $\mathbf{V}^{0}$, $\mathbf{M}_{aux}^{0}$, and receives clients' data $\{\mathcal{D}_{u_{i}}'\}_{u_{i}\in\mathcal{U}}$
      \For {each round t =0, ..., $T-1$}
        \State // execute on server sides
        \State $\mathbf{M}_{aux}^{t+1}$ $\leftarrow$ update model $\mathbf{M}_{aux}^{t}$ on $\{\mathcal{D}_{u_{i}}'\}_{u_{i}\in\mathcal{U}}$
        % \State sample a fraction of clients $\mathcal{U}^{t}$ from $\mathcal{U}$
          \For{$u_{i}\in \mathcal{U}$ \textbf{in parallel}} 
          \State $\widetilde{\mathcal{D}}_{u_{i}}^{t}\leftarrow$ generate augmented data using $\mathbf{M}_{aux}^{t+1}$ with E.q.~\ref{eq_naive_aug}
          \State // execute on client sides
          \State $\mathbf{V}_{u_i}^{t+1}\leftarrow$\Call{ClientTrain}{$\mathbf{V}^{t}$, $\widetilde{\mathcal{D}}_{u_{i}}^{t}$, $\mathbf{U}^{t+1}_{aux}$}
          \EndFor
        \State // execute on central server
        \State $\mathbf{V}^{l+1}\leftarrow$ aggregate received client model parameters $\{\mathbf{V}_{u_i}^{t+1}\}_{u_{i}\in \mathcal{U}}$
        \State update $\mathbf{V}^{l+1}$ using E.q.~\ref{eq_item_con_loss}
      \EndFor
      \Function{ClientTrain} {$\mathbf{V}^{t}$, $\widetilde{\mathcal{D}}_{u_{i}}^{t}$, $\mathbf{U}^{t+1}_{aux}$}
      \State $\mathbf{u}_{i}^{t+1}, \mathbf{V}_{u_i}^{t+1}\leftarrow$ update local model with E.q.~\ref{eq_user_final_loss}
      \State \Return $\mathbf{V}_{u_i}^{t+1}$
      \EndFunction
      \end{algorithmic}
  \end{algorithm}

\section{Experiments}
In this work, we explore the following research questions (RQs):
\begin{itemize}
    \item[$\bullet$] \textbf{RQ1.} How is the performance of \modelname compared to the base FedRecs?
    \item[$\bullet$] \textbf{RQ2.} What is the impact of privacy-preserving mechanisms on data contribution?
    \item[$\bullet$] \textbf{RQ3.} What is the impact of data augmentation module?
    \item[$\bullet$] \textbf{RQ4.} What is the impact of contrastive learning module?
\end{itemize}

\subsection{Datasets}
In this work, we conduct experiments on two widely used recommendation datasets, MovieLens-1M\footnote{\url{https://grouplens.org/datasets/movielens/1m/}} (ML) and Amazon Office Products (AZ)\footnote{\url{https://jmcauley.ucsd.edu/data/amazon/}}, which are collected from different platforms in various scenarios.
ML contains $6,040$ users and $3706$ movies with $1,000,000$ user rating records.
AZ includes $53,258$ reviews among $4905$ users and $2421$ office products.
Following most recommendations with implicit feedback data preprocessing~\cite{he2017neural,zhang2022pipattack,yuan2023interaction}, we convert all interacted items' ratings into $1$ and randomly sample negative items from non-interacted item pools with ratio $1:4$.
Besides, $80\%$ data are used for training while the remaining data are for testing.

\subsection{Evaluation Metrics}
We leverage Recall@K and NDCG@K to evaluate the recommendation performance, and K is set to $20$.
We treat all non-interacted items as the candidate set to avoid sampling-based evaluation bias~\cite{krichene2020sampled}.

\subsection{Baselines}
To show the effectiveness of \modelname, we employ the following baselines:
\begin{itemize}
    \item[$\bullet$] \textbf{FCF}~\cite{ammad2019federated}. It is the first work that combines federated learning with collaborative filtering.
    \item[$\bullet$] \textbf{FedMF}~\cite{chai2020secure}. It combines federated learning with matrix factorization and utilizes a homomorphic encryption technique to ensure user privacy.
    \item[$\bullet$] \textbf{FedNCF}. This is the system that removes the data contribution of \modelname. The details of it are described in Section~\ref{sec_preliminaries}.
    \item[$\bullet$] \textbf{FedNCF+Aug}. We employ privacy-preserving data contribution to FedNCF but only utilize the data with the data augmentation method proposed in Section~\ref{sec_enhance_with_data_contribution}.
    \item[$\bullet$] \textbf{FedNCF+CL}. We employ privacy-preserving data contribution to FedNCF with a contrastive learning method proposed in Section~\ref{sec_enhance_with_data_contribution}. 
\end{itemize}

\subsection{Implementation Details}
Following~\cite{yuan2023hetefedrec,qu2024towards,he2017neural}, the dimensions of user and item embeddings in the NCF model are set to $32$. 
Three feedforward layers with sizes $32$, $16$, and $8$ are used to capture user and item collaborative information.
The maximum number of global training rounds is 20, and all clients are ensured to participate during a global round.
Specifically, at the start of a global round, a shuffled order of clients is generated.
Then, a batch of $256$ clients is sequentially selected to train the model. 
The local training epoch for each client is $5$. 
The optimizer is Adam with $0.001$ learning rate.
The values of $\epsilon$, $\alpha$, $\beta$, and $\lambda$ are set to $5$, $30$, $0.5$, and $0.5$, respectively, and we will explore their influences in the following sections.

\begin{table}[]
    \centering
    \caption{Comparison between \modelname and baselines. ``Aug'' is short for data augmentation and ``CL'' is abbreviated for contrastive learning.}\label{tb_main}
    \resizebox{\textwidth}{!}{
    \begin{tabular}{l|cc|cc}
    \hline
    \multirow{2}{*}{}                  & \multicolumn{2}{c|}{\textbf{ML}}      & \multicolumn{2}{c}{\textbf{AZ}}       \\ \cline{2-5} 
                                       & \textbf{Recall@20} & \textbf{NDCG@20} & \textbf{Recall@20} & \textbf{NDCG@20} \\ \hline
     \textbf{FCF}                      & 0.01286          & 0.02142        & 0.00724        & 0.00326                \\
    \textbf{FedMF}                     & 0.01420          & 0.02273        & 0.00763        & 0.00338                \\
    \textbf{FedNCF}                    & 0.01529          & 0.02355        & 0.00875          & 0.00385                \\ \hline
    \textbf{FedNCF+Aug}                & 0.01580          & 0.02400        & 0.01341          & \textbf{0.00670}                \\
    \textbf{FedNCF+CL}                 & 0.01547          & 0.02377        & 0.00910          & 0.00368                \\
    \textbf{FedNCF+Aug+CL (\modelname)}& \textbf{0.01679}          & \textbf{0.02557}        & \textbf{0.01387}          & 0.00663                \\ \hline
    \end{tabular}}
\end{table}

\subsection{Effectiveness of \modelname (RQ1)}
Table~\ref{tb_main} presents the comparison results of \modelname with baselines on the ML and AZ datasets. Overall, \modelname achieves superior performance among these baselines across most metrics on both datasets. By comparing FedNCF with FedNCF+Aug and FedNCF+CL, we observe that incorporating data contributions through either our proposed augmentation method or the contrastive learning approach enhances the base FedRec performance. Furthermore, the influence of data augmentation is relatively better than that of contrastive learning, indicating that clients in FedRec suffer from a severe data scarcity problem. Finally, the combination of augmentation and contrastive learning, as implemented in \modelname, further improves performance, showcasing its effectiveness. Specifically, \modelname improves FedNCF's Recall@20 scores from $0.0152$ to $0.0167$ on ML and from $0.008$ to $0.0138$ on AZ.

\subsection{Impact of Privacy-preserving data contribution (RQ2)}
\begin{figure}[!t]
    \centering
    \subfloat[Performance Trend on ML with $\epsilon$.]{\includegraphics[width=0.5\textwidth]{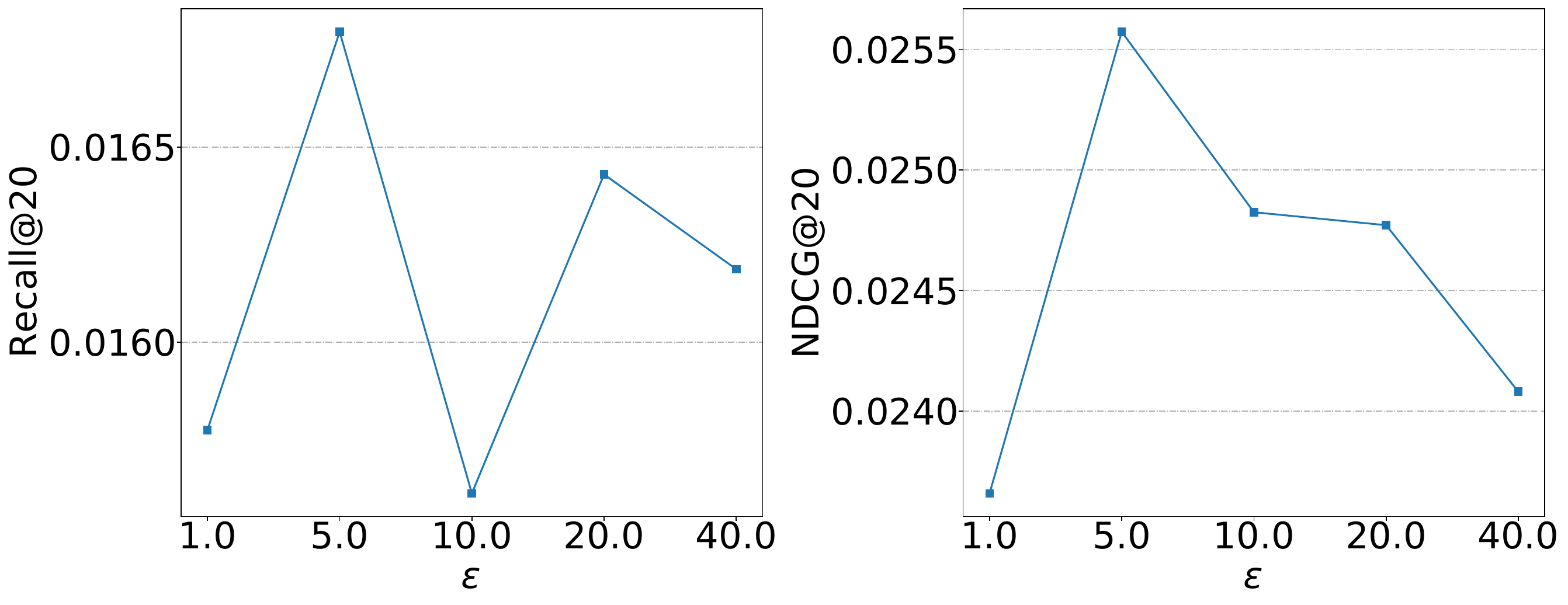}\label{fig_privacy_ml}}
    % \hfil
    \subfloat[Performance Trend on AZ with $\epsilon$.]{\includegraphics[width=0.5\textwidth]{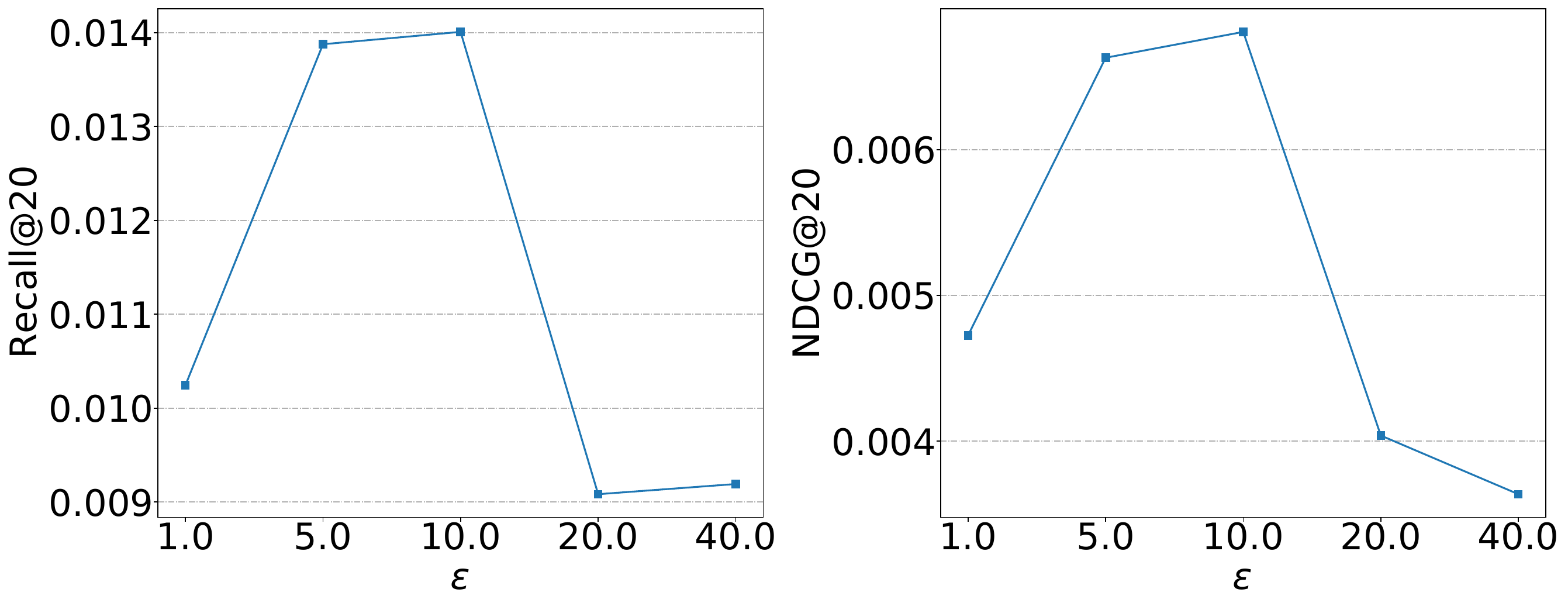}\label{fig_privacy_office}}
    \caption{The performance trend with privacy controller $\epsilon$.}\label{fig_privacy_hyperparameter}
\end{figure}

One major contribution of this paper is the proposal of a privacy-preserving data contribution method. In this section, we investigate the relationship between utility and privacy protection. The privacy-preserving ability of \modelname is controlled by $\epsilon$. Theoretically, privacy protection is negatively correlated with $\epsilon$. In our case, a small $\epsilon$ adds more randomness to item perturbation, while a larger $\epsilon$ makes the exponential mechanism tend to keep the item unchanged.
Figure~\ref{fig_privacy_hyperparameter} shows the performance trends on the ML and AZ datasets. Generally, as $\epsilon$ increases, the model performance first rises to a peak and then gradually declines. This phenomenon is caused by two factors. Firstly, when $\epsilon$ is too small, the uploaded user data becomes random, making it difficult for the system to learn useful information. Secondly, when $\epsilon$ is too large, the uploaded data closely resembles the original dataset. Therefore, when using the auxiliary model as a data augmenter, the generated data will be too similar to the client's original data, resulting in ineffectiveness.

\begin{figure}[!t]
    \centering
    \subfloat[Performance Trend on ML with $\alpha$.]{\includegraphics[width=0.5\textwidth]{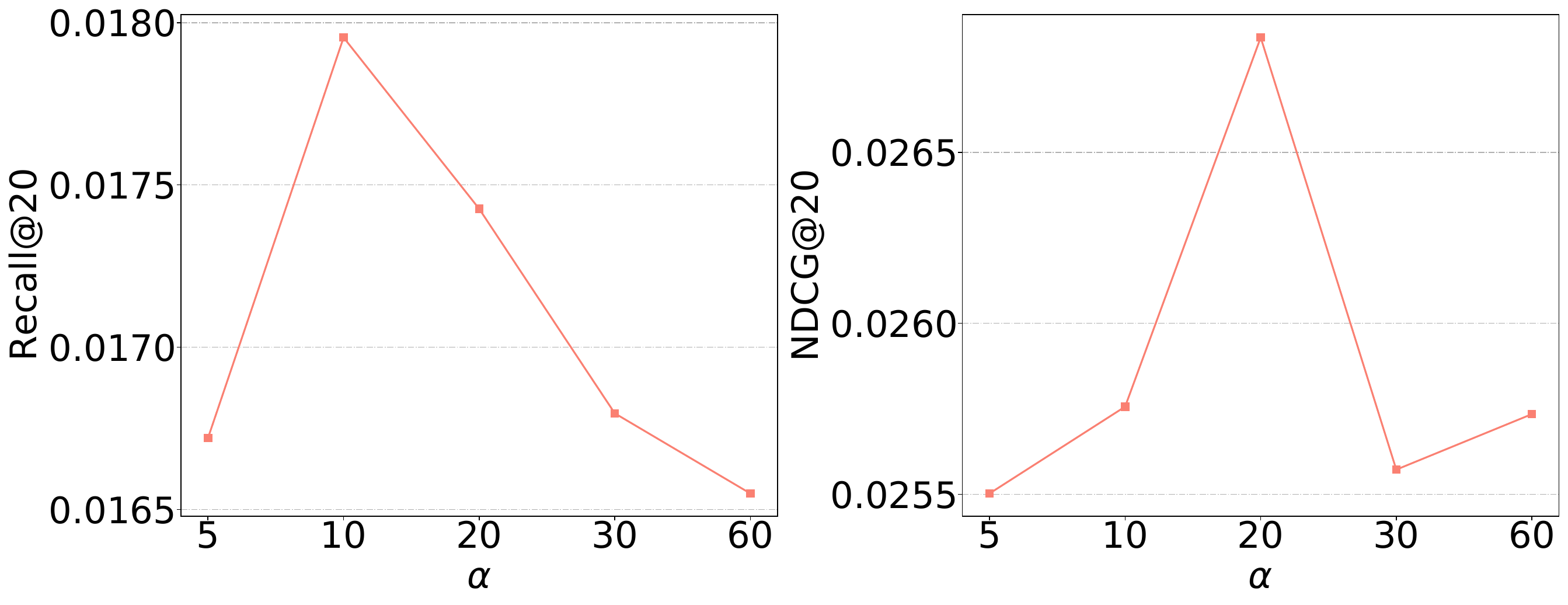}\label{fig_aug_ml}}
    % \hfil
    \subfloat[Performance Trend on AZ with $\alpha$.]{\includegraphics[width=0.5\textwidth]{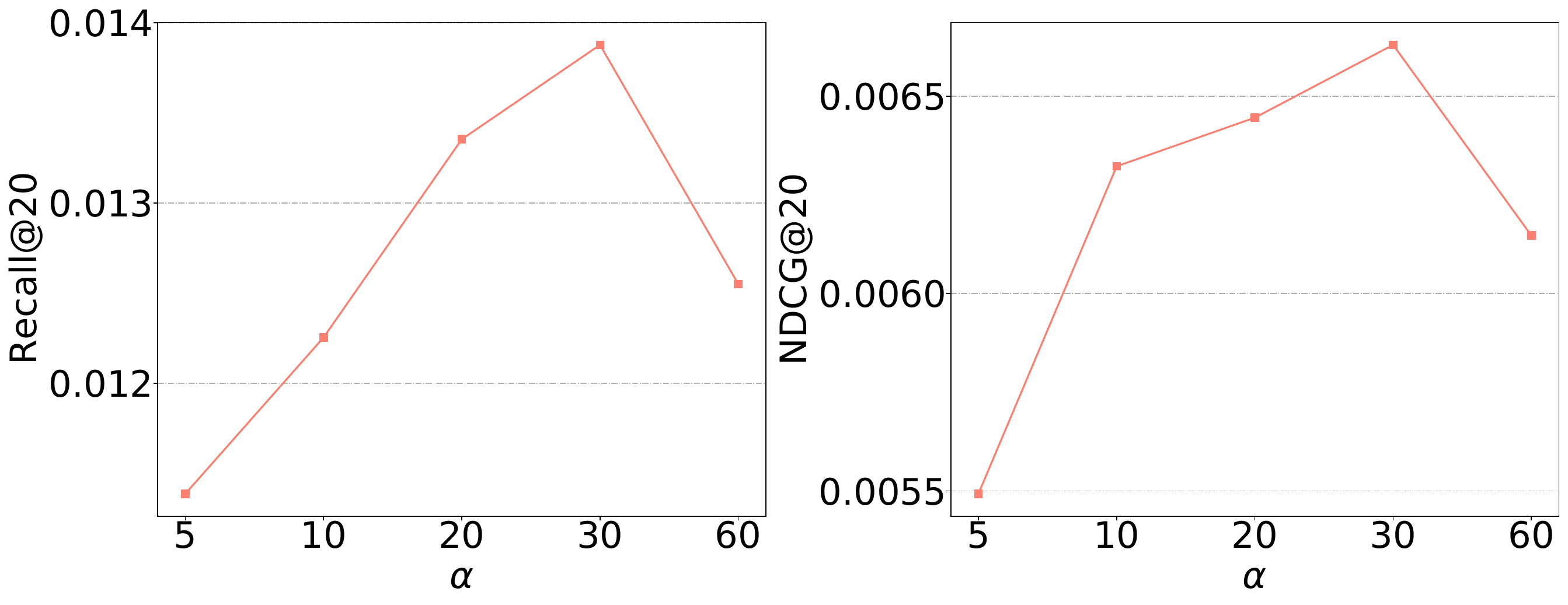}\label{fig_aug_office}}
    \caption{The performance trend with augmentation dataset size $\alpha$.}\label{fig_augment_hyperparameter}
\end{figure}

\subsection{Impact of Data Augmentation (RQ3)}
Utilizing the auxiliary model $\mathbf{M}_{aux}$ to augment clients' local datasets is an effective way to leverage users' noisy data contributions, as demonstrated in Table~\ref{tb_main}. 
In this subsection, we further investigate the influence of the augmented dataset size, $\alpha$, on model performance.
As depicted in Figure~\ref{fig_augment_hyperparameter}, we explore augmented dataset sizes ranging from $5$ to $60$ on the ML and AZ datasets. The performance trends on these two datasets are similar: model performance increases to a peak and then decreases as the augmented dataset size continues to grow. This indicates that an appropriate amount of data augmentation can improve model performance, however, if the augmentation set becomes too large, it will impede the model's ability to learn from the original dataset.

\subsection{Impact of Contrastive Learning (RQ4)}
\begin{figure}[!t]
    \centering
    \subfloat[Performance Trend on ML with $\lambda$.]{\includegraphics[width=0.5\textwidth]{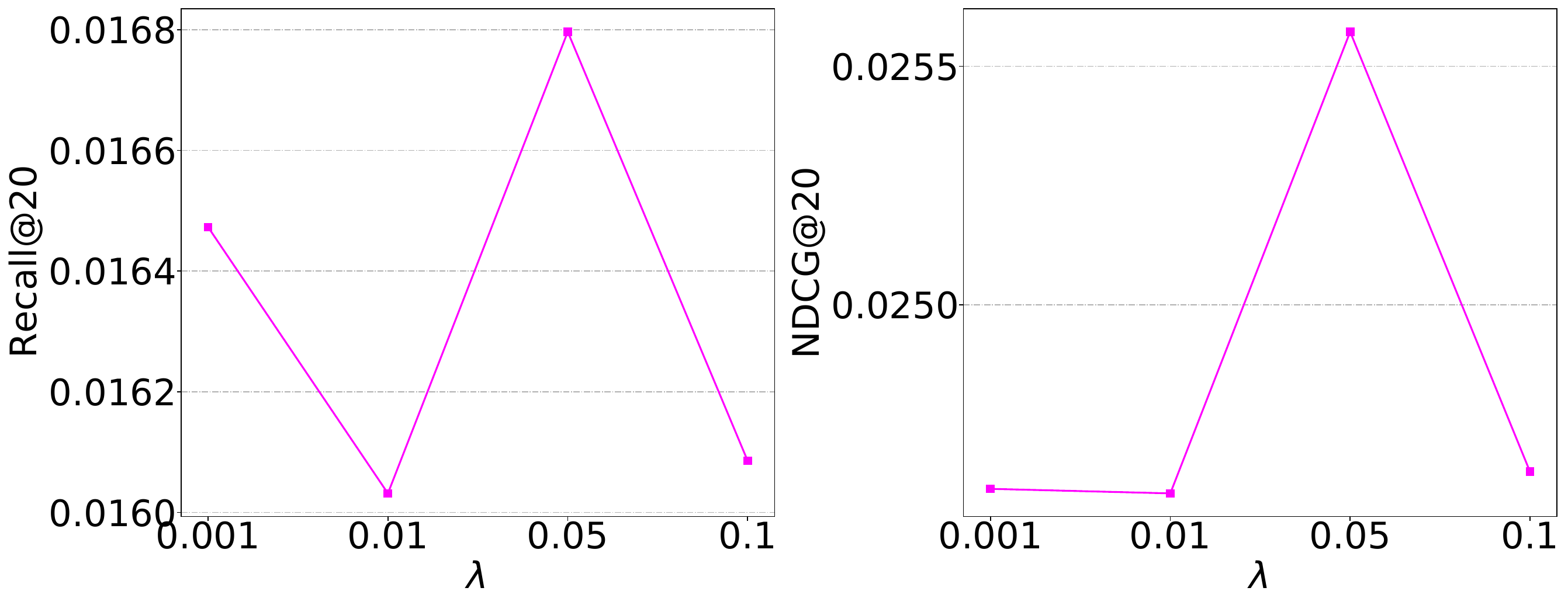}\label{fig_ucl_ml}}
    % \hfil
    \subfloat[Performance Trend on AZ with $\lambda$.]{\includegraphics[width=0.5\textwidth]{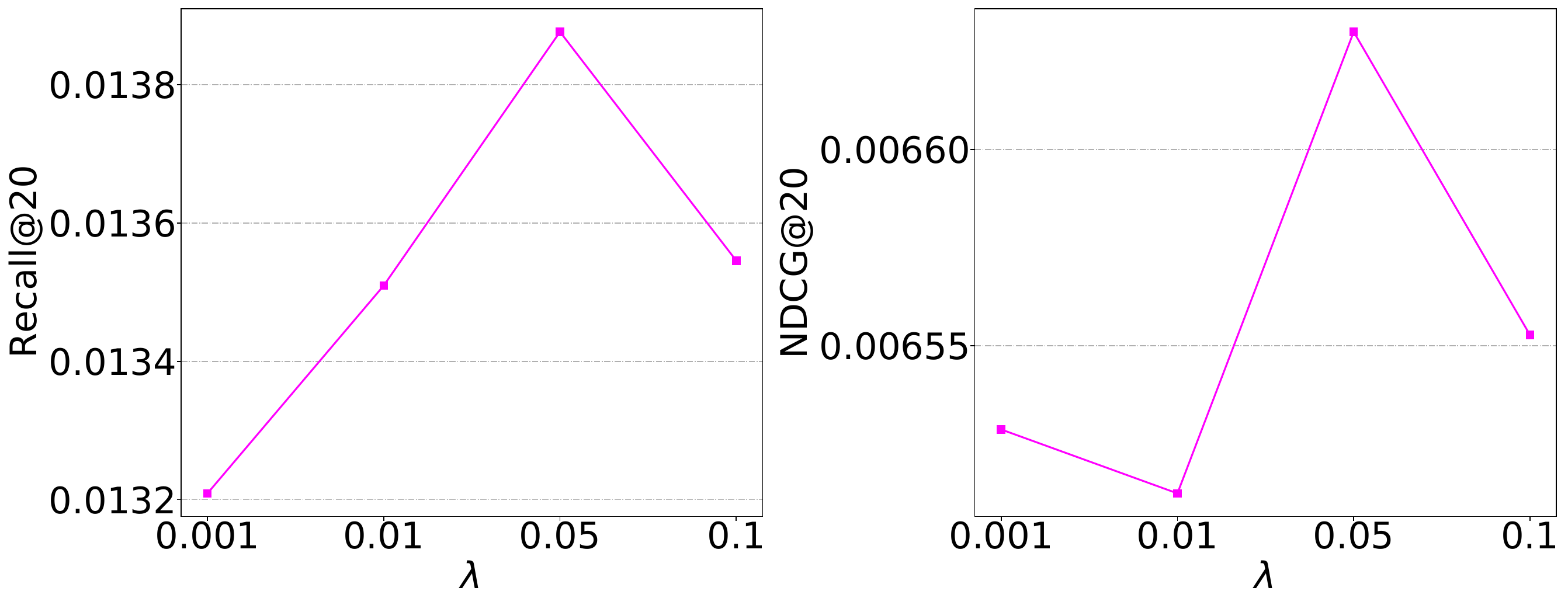}\label{fig_ucl_office}}
    \caption{The performance trend with user contrastive learning $\mathcal{L}^{userCL}$ factor $\lambda$.}\label{fig_ucl_hyperparameter}
\end{figure}

\begin{figure}[!t]
    \centering
    \subfloat[Performance Trend on ML with $\beta$.]{\includegraphics[width=0.5\textwidth]{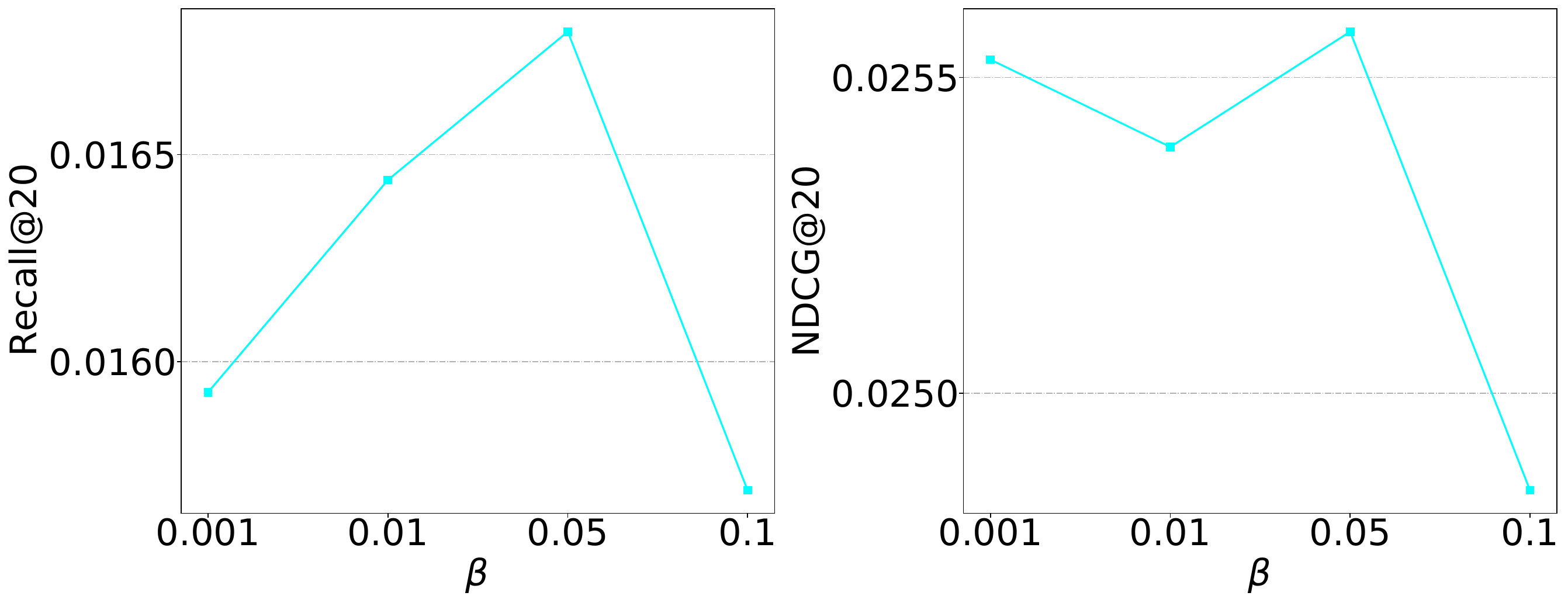}\label{fig_icl_ml}}
    % \hfil
    \subfloat[Performance Trend on AZ with $\beta$.]{\includegraphics[width=0.5\textwidth]{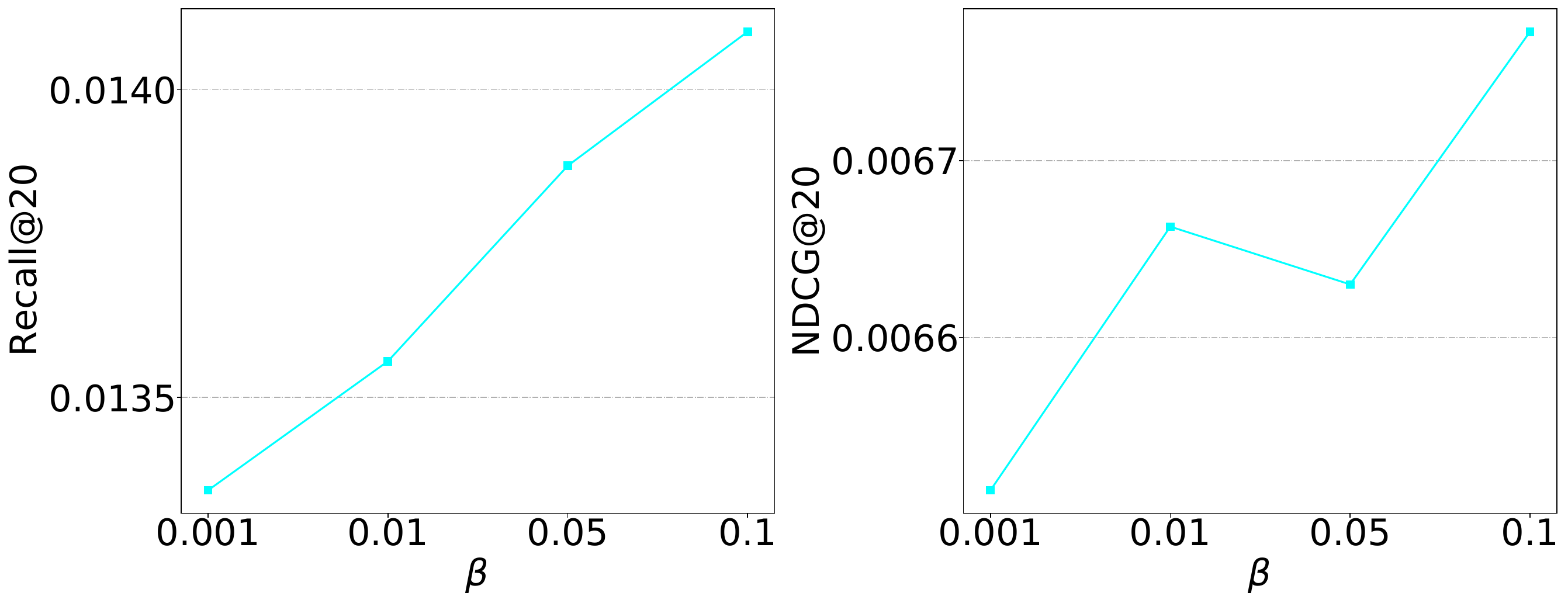}\label{fig_icl_office}}
    \caption{The performance trend with item contrastive learning $\mathcal{L}^{itemCL}$ factor $\beta$.}\label{fig_icl_hyperparameter}
\end{figure}

Treating the auxiliary model $\mathbf{M}_{aux}$ as a contrastive learning view is a novel method to leverage user data contributions. \modelname includes two contrastive learning tasks, $\mathcal{L}^{userCL}$ and $\mathcal{L}^{itemCL}$, corresponding to user and item embeddings, respectively. These two tasks are controlled by factors $\lambda$ and $\beta$.
Figure~\ref{fig_ucl_hyperparameter} shows the performance changes with varying $\lambda$. \modelname achieves the best performance when $\lambda=0.05$ on both the ML and AZ datasets. Figure~\ref{fig_icl_hyperparameter} presents the impact of $\beta$. Similar to $\lambda$, on the ML dataset, \modelname obtains the best performance when $\beta=0.05$, while on the AZ dataset, the best performance is achieved when $\beta=0.1$.
Overall, based on these two figures, we find that too small factors limit the influence of contrastive learning, while too large factors make contrastive learning overwhelming, impeding the model learning process.

\section{Conclusion and Future Work}
In this paper, we propose a novel federated recommendation framework, \modelname, which enables users to contribute their data to enhance FedRec performance in a privacy-preserving manner. Specifically, \modelname employs the exponential mechanism to protect users' data before uploading it to the central server. To effectively utilize these noisy data contributions, an auxiliary model is trained upon them, acting as a data augmentor and providing contrastive learning views during FedRec's training. Extensive experiments on two datasets demonstrate the effectiveness of \modelname.

\bibliographystyle{splncs04}
\bibliography{samplepaper} 
%
% \begin{thebibliography}{8}
% \bibitem{ref_article1}
% Author, F.: Article title. Journal \textbf{2}(5), 99--110 (2016)

% \bibitem{ref_lncs1}
% Author, F., Author, S.: Title of a proceedings paper. In: Editor,
% F., Editor, S. (eds.) CONFERENCE 2016, LNCS, vol. 9999, pp. 1--13.
% Springer, Heidelberg (2016). \doi{10.10007/1234567890}

% \bibitem{ref_book1}
% Author, F., Author, S., Author, T.: Book title. 2nd edn. Publisher,
% Location (1999)

% \bibitem{ref_proc1}
% Author, A.-B.: Contribution title. In: 9th International Proceedings
% on Proceedings, pp. 1--2. Publisher, Location (2010)

% \bibitem{ref_url1}
% LNCS Homepage, \url{http://www.springer.com/lncs}, last accessed 2023/10/25
% \end{thebibliography}
\end{document}